\documentclass[journal]{IEEEtran}

\usepackage{amsmath,amssymb,amsfonts}
\usepackage{bm}
\usepackage{graphicx}
\usepackage{booktabs}
\usepackage{multirow}
\usepackage{xcolor}
\usepackage{cite}
\usepackage{url}
\usepackage{hyperref}

\title{Radar-Aided Near-Field Beam Prediction via Beam Map Learning for XL-MIMO V2I Communications}

\author{Jiali~Nie,~\IEEEmembership{Student Member,~IEEE,}
Yu~Han,~\IEEEmembership{Member,~IEEE,}
Yuanhao~Cui,~\IEEEmembership{Member,~IEEE,}
Xiaojie~Li,~\IEEEmembership{Student Member,~IEEE,}
Shi~Jin,~\IEEEmembership{Fellow,~IEEE,}
and~Chao-Kai~Wen,~\IEEEmembership{Fellow,~IEEE}
\thanks{Received xxx; revised xxx; accepted xx. Date of publication xx.
This work was supported in part by the xxxx. The work of Jiali Nie was supported in part by the SEU Innovation Capability Enhancement Plan for Doctoral Students CXJH SEU 26104. The code for this paper is available at \url{https://github.com/fly-winder/Radar2BeamMap-NF} (Corresponding author: Yu Han; Shi
Jin.)}
\thanks{Jiali Nie, Yu Han, Xiaojie Li, and Shi Jin are with the School of Information Science and Engineering, Southeast University, Nanjing, China (e-mail: \{niejl, hanyu, xiaojieli, jinshi\}@seu.edu.cn).}
\thanks{Yuanhao Cui is with the School of Information and Communication Engineering, Beijing University of Posts and Telecommunications, Beijing, China (e-mail: cuiyuanhao@bupt.edu.cn).}
\thanks{Chao-Kai Wen is with the Institute of Communications Engineering, National Sun Yat-sen University, Kaohsiung, Taiwan (e-mail: chaokai.wen@mail.nsysu.edu.tw).}
}

\begin{document}
\markboth{IEEE xxxx,~Vol.~XX, No.~XX}%
{Nie \MakeLowercase{\textit{et al.}}: Radar-Aided Near-Field Beam Prediction via Beam Map Learning}
\maketitle

\begin{abstract}
Near-field beam training in extremely large-scale multiple-input multiple-output (XL-MIMO) vehicle-to-infrastructure (V2I) systems incurs high overhead due to large range-angle codebooks and rapid channel variation. This paper proposes a passive radar-aided framework for near-field beam prediction based on radar-to-beam map learning. By exploiting the spatial correlation between radar observations and communication signals, the proposed method maps radar Bartlett spectra to communication beam maps using a lightweight encoder-decoder convolutional neural network. Gaussian soft supervision is further introduced to preserve beam-space continuity. Simulations on a synchronized Sionna ray tracing radar-communication dataset show that the proposed method consistently improves Top-k accuracy, distance-based accuracy, beam loss, and spectral efficiency.
\end{abstract}


\section{Introduction}
Future vehicle-to-infrastructure (V2I) communication systems require high-rate and low-latency links for connected and automated vehicles. To provide sufficient link budget at high carrier frequencies, large-scale and extremely large-scale multiple-input multiple-output (XL-MIMO) arrays are expected to be deployed at roadside units (RSUs)~\cite{10379539}. Such arrays provide highly directional beams and large beamforming gains, but also make beam training increasingly challenging. As the array aperture increases, vehicles within practical V2I ranges may lie in the radiating near-field region, where the beam response becomes dependent on both angle and range~\cite{cui2023nearfield}. Therefore, near-field beam selection requires searching over a large range-azimuth-elevation codebook, which is prohibitive for highly dynamic V2I links with frequent beam updates~\cite{nie2025nearfield}.

Integrated sensing and communications (ISAC) provides a promising paradigm for reducing beam training overhead by exploiting environmental sensing information~\cite{liu2022isac}. Existing sensing-aided beam prediction studies have used LiDAR point clouds, visual images, and location information to infer candidate beam directions~\cite{cui2024sensing,jiang2023lidar,charan2025sensing}. These methods demonstrate the potential of multimodal sensing for beam management, but they mainly rely on geometric cues or dominant LoS-related features, and their robustness may degrade under dynamic blockage or NLoS propagation.

Radar sensing is attractive for V2I beam management since automotive radars provide high-resolution range and angular information about vehicles and scatterers~\cite{liu2020radar}. Existing radar-aided methods reduce beam search overhead using target estimates, covariance matrices, angular spectra, or deep learning-based mappings from radar observations to mmWave/sub-THz beam indices~\cite{demirhan2022radar,demirhan2026radartracking}. Passive radar at the RSU has also been exploited to configure V2I links by intercepting automotive radar signals~\cite{ali2020passive}, while low-dimensional radar spatial features have been mapped to communication beam representations~\cite{graff2023deep}.  Nevertheless, most existing methods assume far-field propagation with uniform linear arrays or low-dimensional angular codebooks. In XL-MIMO V2I systems, the large array aperture makes the beam response range-angle dependent, requiring beam training over a three-dimensional range-azimuth-elevation codebook. This greatly increases training overhead, while high vehicle mobility further aggravates beam misalignment. In addition, hard beam-index classification ignores beam-space continuity, and its robustness under dynamic blockage and corner NLoS propagation remains insufficiently studied.

In this paper, we propose a radar-aided near-field beam prediction framework based on radar-to-beam map learning. Instead of directly predicting a hard beam index, the proposed method maps radar Bartlett spectra to communication angular beam maps using a lightweight encoder-decoder CNN. A Gaussian soft label and a neighborhood-weighted mean squared error (MSE) loss are designed to preserve beam-space continuity. Experiments on a synchronized Sionna RT-based radar-communication dataset covering LoS, dynamic blockage, and corner NLoS scenarios demonstrate superior beam prediction and communication performance.

\section{System Model}
\begin{figure}[t]
    \centering
    \includegraphics[width=0.4\textwidth]{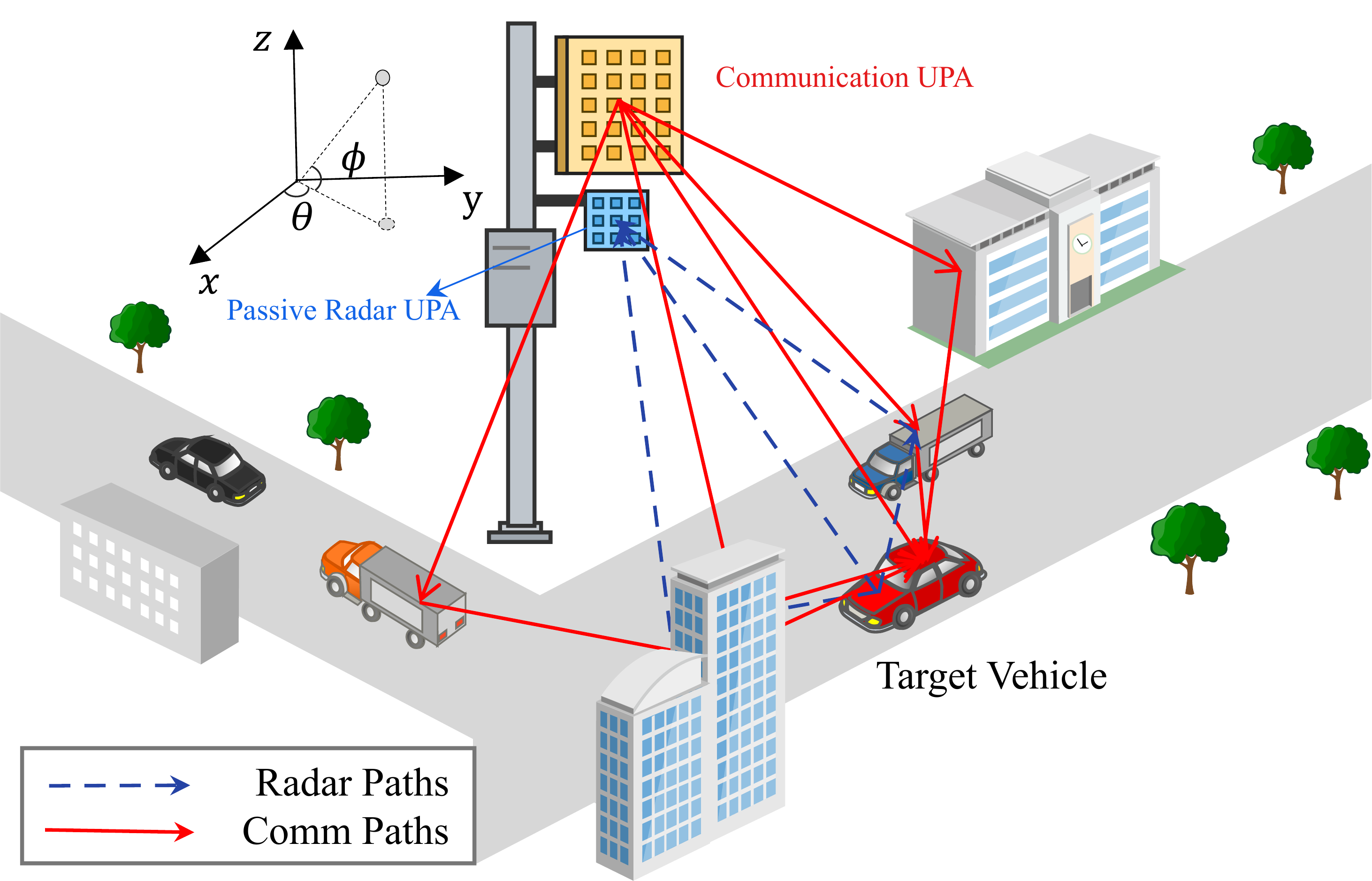}
    \caption{System model of the passive radar-aided XL-MIMO V2I communications.}
    \label{fig:system_model}
\end{figure}
We consider a V2I downlink system, as shown in Fig.~\ref{fig:system_model}, where an ISAC-enabled roadside XL-MIMO base station (BS) serves a highly mobile vehicle. The roadside BS is equipped with a large-scale uniform planar array (UPA) operating in the upper 6 GHz (U6G) band for wireless communication, together with a passive sensing array for intercepting frequency-modulated continuous-wave (FMCW) signals transmitted by the onboard automotive radar. The ego vehicle is equipped with a single communication antenna and an FMCW radar, which are spatially separated.

\subsection{Communication Model}
Consider a V2I downlink where the roadside BS employs a UPA with $N_c=N_{c,v}N_{c,h}$ elements on the $y$-$z$ plane at height $H$. The array center is $\mathbf{p}_{\mathrm{BS}}=[0,0,H]^T$, and the $(m,n)$-th element has position $\bar{\mathbf{p}}_{m,n}$, where $m=0,\ldots,N_{c,v}-1$ and $n=0,\ldots,N_{c,h}-1$. For a point located at spherical coordinates $(r,\theta,\phi)$ where $r$, $\theta$, and $\phi$ denote the range, azimuth angle, and elevation angle,
respectively, the direction vector is 
\begin{equation}
\mathbf{u}(\theta,\phi)
=
[\cos\phi\cos\theta,\,
\cos\phi\sin\theta,\,
\sin\phi]^T.
\label{eq:direction_vector}
\end{equation}

Due to the large XL-MIMO aperture, the vehicle may lie in the radiating near-field region. The spherical-wave near-field steering vector $\mathbf{a}_{\mathrm{NF}}(r,\theta,\phi)\in\mathbb{C}^{N_c\times 1}$ is
\begin{equation}
\left[
\mathbf{a}_{\mathrm{NF}}(r,\theta,\phi)
\right]_{m,n}
=
\frac{1}{\sqrt{N_c}}
e^{-j\frac{2\pi}{\lambda_c}
\left(
\left\|
r\mathbf{u}(\theta,\phi)
-
\bar{\mathbf{p}}_{m,n}
\right\|
-r
\right)},
\label{eq:near_field_steering}
\end{equation}
where $\lambda_c$ is the communication wavelength. For a wideband OFDM system with $K$ subcarriers and subcarrier spacing $\Delta f$, the near-field channel at the $k$-th subcarrier is modeled as
\begin{equation}
\mathbf{h}_{k,t}
=
\sum_{\ell=1}^{L_t}
\alpha_{\ell,t}
e^{j2\pi \nu_{\ell,t}t}
e^{-j2\pi k\Delta f \tau_{\ell,t}}
\mathbf{a}_{\mathrm{NF}}
(r_{\ell,t},\theta_{\ell,t},\phi_{\ell,t}),
\label{eq}
\end{equation}
where $L_t$ is the number of communication paths at time $t$, $\alpha_{\ell,t}$, $\tau_{\ell,t}$, $\nu_{\ell,t}$, and
$(r_{\ell,t},\theta_{\ell,t},\phi_{\ell,t})$ denote the complex gain,
delay, Doppler shift, and range-azimuth-elevation parameters of the
$\ell$-th communication path at time $t$, respectively. Given a near-field beamforming vector $\mathbf{w}\in\mathbb{C}^{N_c\times 1}$, the received SNR at the $k$-th subcarrier is
\begin{equation}
\gamma_{k,t}(\mathbf{w})
=
\frac{
P_t
\left|
\mathbf{h}_{k,t}^{H}\mathbf{w}
\right|^2
}{
\sigma_n^2
},
\label{eq:snr}
\end{equation}
where $P_t$ is the transmit power and $\sigma_n^2$ is the noise power. The beamforming vector is selected from a predefined near-field codebook to provide sufficient array gain for the highly directional V2I link.

\subsection{Radar Model}
To provide environmental priors for V2I networking, a passive sensing array with $N_r=N_{r,v}N_{r,h}$ antenna elements is deployed at the roadside BS to intercept onboard automotive radar signals. For simplicity, only the target vehicle is considered as the opportunistic radar transmitter. The array is placed on the $y$-$z$ plane with inter-element spacing $d_r=\lambda_r/2$, where $\lambda_r$ is the radar wavelength. The vehicle radar operates at millimeter wave and transmits FMCW chirps. During one chirp duration $T_p$, the transmitted signal is
\begin{equation}
s_r(t_c)
=
\sqrt{P_r}
e^{j2\pi
\left(
f_r t_c+
\frac{\beta t_c^2}{2}
\right)
+j\psi},
\quad
t_c\in[0,T_p],
\label{eq:fmcw_signal}
\end{equation}
where $P_r$ is the radar transmit power, $f_r$ is the starting frequency, $\beta=B_r/T_p$ is the chirp rate, $B_r$ is the sweep bandwidth, and $\psi$ is the initial phase. The received radar signal at the roadside BS is modeled as
\begin{equation}
\mathbf{y}_r(t_c)
=
\sum_{\ell=1}^{L_r}
\alpha_{r,\ell}
\mathbf{a}_{r}(\theta_{r,\ell},\phi_{r,\ell})
s_r(t_c-\tau_{r,\ell})
+
\mathbf{n}_r(t_c),
\label{eq:radar_received_signal}
\end{equation}
where $L_r$ denotes the number of radar-visible paths associated with a subset of the communication channel. Moreover, $\alpha_{r,\ell}$, $\tau_{r,\ell}$, and $\mathbf{a}_{r}(\theta_{r,\ell},\phi_{r,\ell})$ denote the complex gain, delay, and radar steering vector of the $\ell$-th radar path, respectively. Since the radar aperture is small relative to the target range, $\mathbf{a}_r(\theta,\phi)\in\mathbb{C}^{N_r\times 1}$ follows the far-field plane-wave assumption.

Following existing passive FMCW radar processing, synchronization and dechirping are assumed to have been completed. After sampling, $N_s$ fast-time samples per chirp and $N_d$ chirps per coherent processing interval (CPI) are collected into a radar cube:
\begin{equation}
\mathbf{X}_{\mathrm{r}}
\in
\mathbb{C}^{N_d\times N_r\times N_s},
\label{eq:radar_cube}
\end{equation}
where the three dimensions correspond to chirps, radar antennas, and fast-time samples, respectively.

\subsection{Problem Formulation}
The roadside BS serves the vehicle using a predefined near-field codebook, which jointly discretizes range, azimuth, and elevation to account for spherical wavefronts:
\begin{equation}
\begin{aligned}
\mathcal{W}
=
\{&
\mathbf{w} (r_q,\theta_i,\phi_j)
\mid
q=1,\ldots,N_{\rho} ,\\
&
i=1,\ldots,N_{\theta} ,
\;
j=1,\ldots,N_{\phi} 
\},
\end{aligned}
\label{eq:fine_codebook}
\end{equation}
where $\mathbf{w} (r_q,\theta_i,\phi_j)$ is a near-field steering vector focused on a range-azimuth-elevation grid point, and $N_{\rho} $, $N_{\theta} $, and $N_{\phi} $ denote the codebook sizes along the three dimensions. Specifically, the range grid is uniformly sampled in the inverse-range domain, i.e., $1/r_q$ is uniformly distributed over $[1/r_{\max},1/r_{\min}]$. The azimuth and elevation grids are uniformly sampled in the spatial-frequency domains $\sin\theta$ and $\sin\phi$, respectively, with $\theta_i\in[\theta_{\min},\theta_{\max}]$ and $\phi_j\in[\phi_{\min},\phi_{\max}]$. The optimal codeword is obtained by maximizing the average beamforming gain over all subcarriers:
\begin{equation}
(q_t^\star,i_t^\star,j_t^\star)
=
\arg\max_{q,i,j}
\frac{1}{K}
\sum_{k=0}^{K-1}
\left|
\mathbf{h}_{k,t}^H
\mathbf{w} (r_q,\theta_i,\phi_j)
\right|^2 .
\label{eq:optimal_codeword}
\end{equation}

The exhaustive search in \eqref{eq:optimal_codeword} provides the ground-truth near-field beam index for supervised learning, but its codebook size $|\mathcal{W}|=N_{\rho} N_{\theta} N_{\phi} $ leads to prohibitive training overhead in high-mobility XL-MIMO V2I systems.

To reduce this overhead, we exploit radar observations to infer the near-field beam index. Although radar and communication links share common structures and scatterers, mismatches in carrier frequency, array aperture, antenna location, and propagation response prevent direct geometric conversion from radar observations to the optimal communication beam. We therefore formulate radar-aided near-field beam selection as a learning-based spatial mapping problem:
\begin{equation}
(\hat q,\hat i,\hat j)
=
\mathcal{G}_{\Theta}(\mathbf{X}_{\mathrm{r}}),
\label{eq:general_mapping}
\end{equation}
where $\mathcal{G}_{\Theta}(\cdot)$ maps the radar observation $\mathbf{X}_{\mathrm{r}}$ to the predicted near-field beam index, so that $\mathbf{w} (r_{\hat q},\theta_{\hat i},\phi_{\hat j})$ approaches the exhaustive-search codeword in \eqref{eq:optimal_codeword}. The detailed radar-to-beam mapping is presented in Section~\ref{sec:proposed_method}.

\begin{figure}[t]
    \centering
    \includegraphics[width=0.75\linewidth]{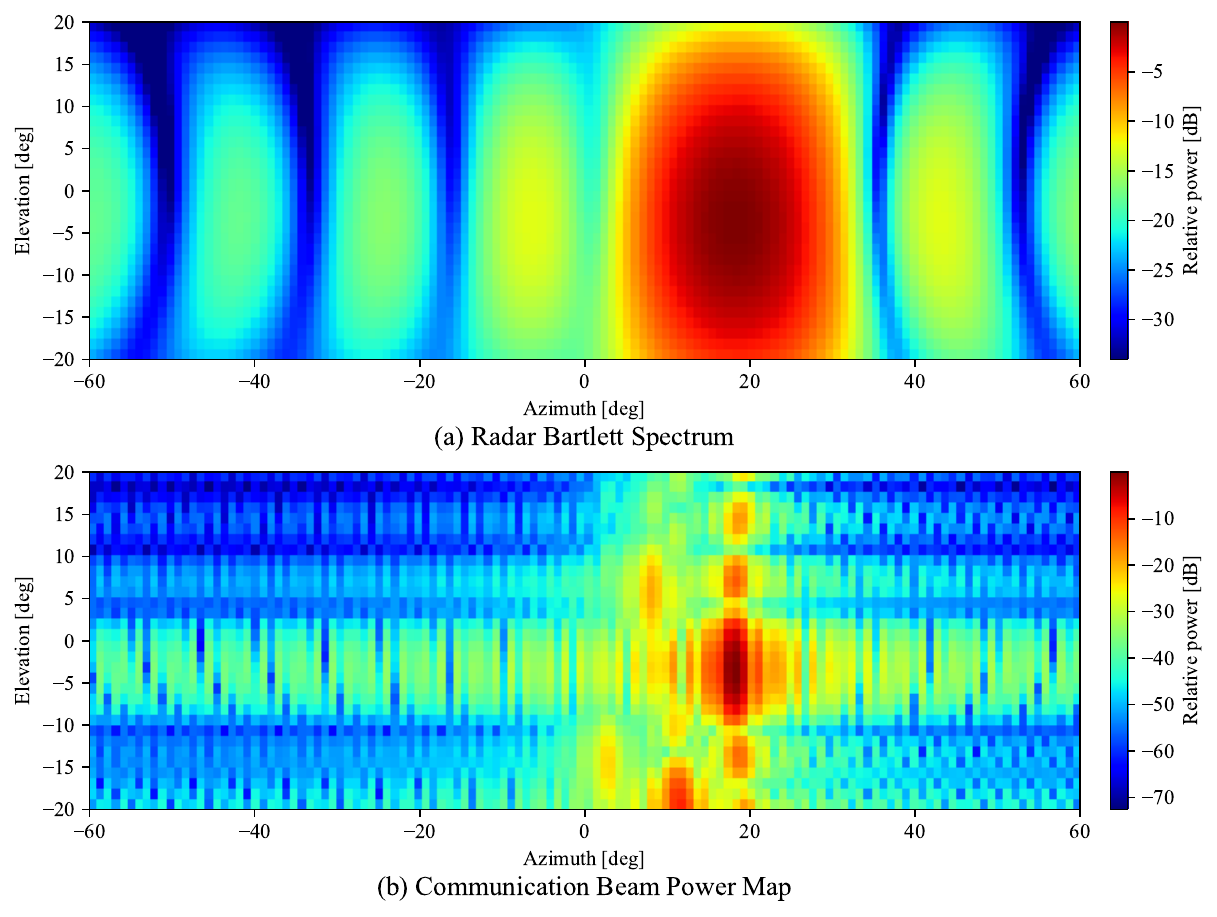}
    \caption{Radar-communication spatial correlation in V2I scenarios}
    \label{fig:radar_comm_correlation}
\end{figure}

\section{Proposed Radar-aided Near-Field Beam Prediction Method}
\label{sec:proposed_method}

\begin{figure*}[t]
    \centering
    \includegraphics[width=0.7\linewidth]{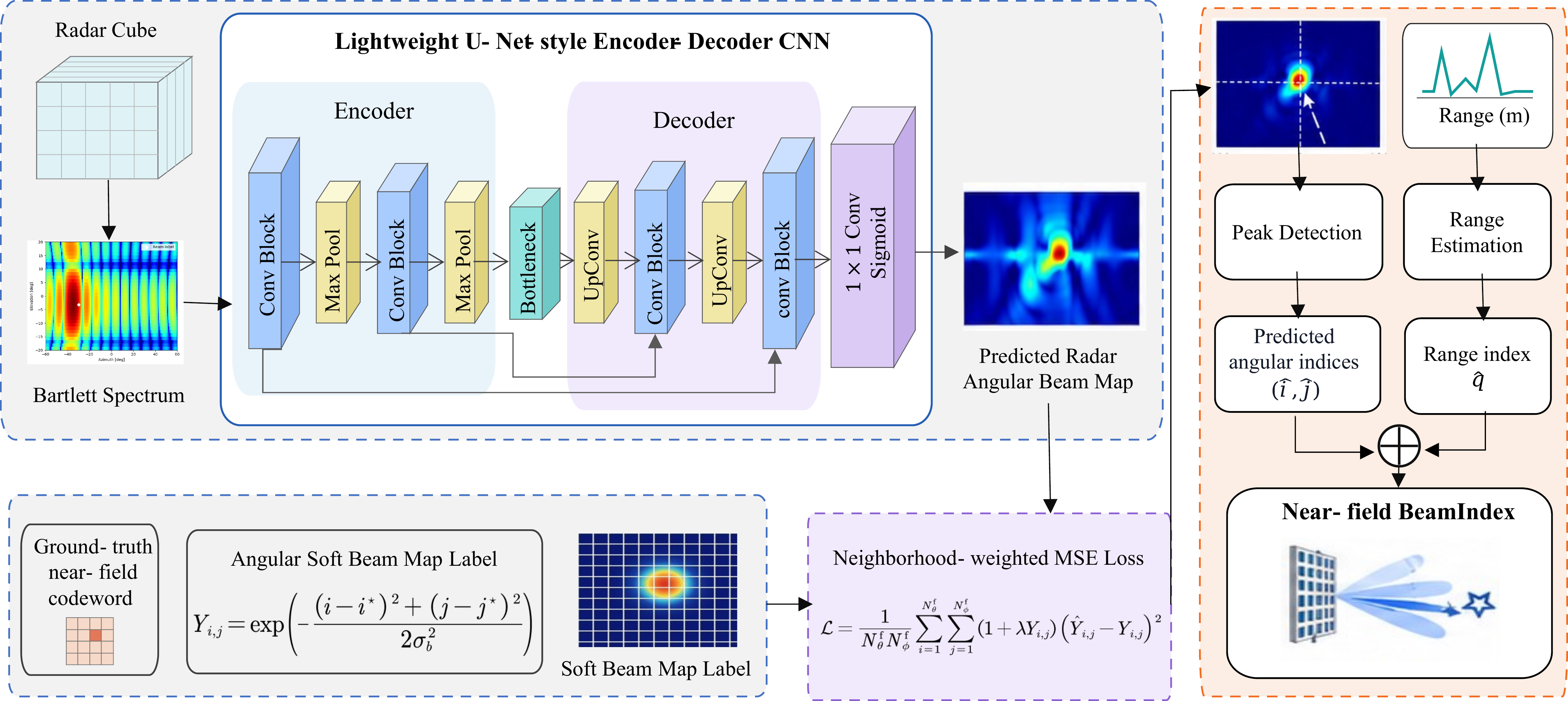}
    \caption{Overall framework of the proposed radar-to-beam map learning method.}
    \label{fig:proposed_framework}
\end{figure*}

\subsection{Motivation and Framework Overview}
From the radar cube, we construct $T$ spatial snapshots stacked in $\mathbf{S}_{\mathrm r}$ and estimate the radar covariance matrix as
\begin{equation}
\mathbf{R}_{\mathrm r}
=
\frac{1}{T}
\mathbf{S}_{\mathrm r}
\mathbf{S}_{\mathrm r}^{H}.
\label{eq:radar_covariance}
\end{equation}

The radar angular representation is then obtained using the Bartlett spectrum
\begin{equation}
\mathbf{P}_{\mathrm r}(\theta,\phi)
=
\mathbf{a}_{\mathrm r}^{H}(\theta,\phi)
\mathbf{R}_{\mathrm r}
\mathbf{a}_{\mathrm r}(\theta,\phi).
\label{eq:bartlett_spectrum}
\end{equation}

As illustrated in Fig.~\ref{fig:radar_comm_correlation}, although $\mathbf{P}_{\mathrm r}(\theta,\phi)$ and the communication beam power map exhibit similar dominant angular structures, their peaks are not always aligned due to radar-communication mismatch. This motivates a learning-based radar-to-beam mapping instead of direct geometric projection.

For the considered array, the Rayleigh distance is about $90$ m, indicating non-negligible near-field effects. Since the considered XL-MIMO beam response is more sensitive to angular mismatch than to small range deviations under the adopted deployment geometry, we use radar ranging to determine the range index and learn only the angular beam map.

For each training sample, the ground-truth near-field codeword $(q^\star,i^\star,j^\star)$ is obtained by exhaustive search. Since the proposed network predicts the angular beam map, only $(i^\star,j^\star)$ is used for supervision. A one-hot label can represent the optimal angular beam, but it ignores the physical continuity of the angular codebook, where adjacent narrow beams often correspond to neighboring directions and similar beamforming gains. To exploit this continuity, we construct a two-dimensional Gaussian soft beam map centered at the optimal angular indices:
\begin{equation}
Y_{i,j}
=
\exp
\left(
-\frac{
(i-i^\star)^2+(j-j^\star)^2
}{
2\sigma_b^2
}
\right),
\label{eq:soft_label}
\end{equation}
where $\sigma_b$ controls the spatial spread over the angular codebook. The beam map reaches its maximum at $(i^\star,j^\star)$ and smoothly decays for neighboring beams. This encoding converts hard beam-index classification into structured beam map regression, better matching codebook beam selection with strong spatial correlation among adjacent beams.

As shown in Fig.~\ref{fig:proposed_framework}, the radar Bartlett spectrum $\mathbf{P}_{\mathrm r}$ is fed into a neural network to predict the angular beam map:
\begin{equation}
\hat{\mathbf{Y}}
=
f_{\Theta}(\mathbf{P}_{\mathrm r}),
\quad
\hat{\mathbf{Y}}
\in
[0,1]^{N_{\theta}\times N_{\phi}},
\label{eq:beamma_prediction}
\end{equation}
where $f_{\Theta}(\cdot)$ denotes the beam map prediction network. The angular indices $(\hat i,\hat j)$ are obtained from the beam map peak and combined with the radar-estimated range index $\hat q$ for final near-field beam selection.

\subsection{Lightweight Beam Prediction Framework}
We adopt a lightweight U-Net-style encoder-decoder CNN to learn the radar-to-beam map from the radar Bartlett spectrum $\mathbf{P}_{\mathrm r}$ to the communication beam map $\hat{\mathbf{Y}}$. As shown in Fig.~\ref{fig:proposed_framework}, the network consists of an encoder, a bottleneck, and a decoder. The encoder uses two convolutional stages with max-pooling to extract multi-scale spatial features, where each stage contains two $3\times3$ convolutional layers followed by batch normalization and ReLU activation. The bottleneck captures high-level spatial representations, while the decoder upsamples the features using transposed convolutions and skip connections to preserve fine angular structures. A final $1\times1$ convolutional layer with sigmoid activation produces the predicted beam map.

The network is trained to minimize the discrepancy between the predicted beam map $\hat{\mathbf{Y}}$ and the soft label $\mathbf{Y}$. Since most beam map entries are close to zero, standard mean squared error may under-emphasize the optimal-beam neighborhood. We therefore adopt a neighborhood-weighted mean squared error loss:
\begin{equation}
\mathcal{L}
=
\frac{1}{N_{\theta}N_{\phi}}
\sum_{i=1}^{N_{\theta}}
\sum_{j=1}^{N_{\phi}}
(1+\lambda Y_{i,j})
\left(
\hat Y_{i,j}
-
Y_{i,j}
\right)^2,
\label{eq:weighted_mse}
\end{equation}
where $\lambda$ controls the emphasis on the optimal-beam neighborhood. This weighting encourages a sharper response around the optimal beam while retaining smooth supervision for adjacent beams. After training, the angular indices $(\hat i,\hat j)$ are obtained from the beam map peak and combined with the radar-estimated range index $\hat q$ to select the final near-field beam $(\hat q,\hat i,\hat j)$. Thus, the proposed method avoids exhaustive scanning of the full three-dimensional near-field codebook by decomposing beam selection into radar-assisted range indexing and learned angular beam map prediction.

\section{Experimental Setup and Result Analysis}

\subsection{Experimental Setup}
We evaluate the proposed method on a Sionna ray-tracing-based synchronized radar-communication dataset with $20$ urban scenarios and a frame interval of $100$ ms. The dataset covers LoS, dynamic blockage, and corner NLoS V2I conditions, and records the radar cube, OFDM channel, near-field beam label, Top-5 beams, received powers, and LoS indicator for each frame. The dataset description is available at \url{https://fly-winder.github.io/SenseComm/}.

For cross-scenario evaluation, samples from $17$ scenarios are split into training and validation sets with an $80\%:20\%$ ratio, while three Shanghai scenarios with $2004$ samples are used only for testing. The network is implemented in PyTorch and trained with Adam. Key simulation and training parameters are listed in Table~\ref{tab:simulation_parameters}.


\begin{table}[t]
\centering
\caption{Simulation and Training Parameters}
\label{tab:simulation_parameters}
\renewcommand{\arraystretch}{1.05}
\setlength{\tabcolsep}{3.0pt}
\begin{tabular}{lclc}
\hline
\textbf{Parameter} & \textbf{Value} & \textbf{Parameter} & \textbf{Value} \\
\hline
Comm. carrier & $7$ GHz & Radar carrier & $77$ GHz \\
Comm. bandwidth & $20$ MHz & Radar bandwidth & $200$ MHz \\
OFDM subcarriers & $K=128$ & Radar SNR & $0$ dB \\
Comm. array & $16\times64$ UPA & Radar array & $4\times8$ UPA \\
Roadside BS height & $10$ m & Road config. & Six-lane \\
$[r_{\min},r_{\max}]$ & $20,100$ m & $N_\rho\times N_\theta\times N_\phi$ & $16\times128\times32$ \\
$[\theta_{\min},\theta_{\max}]$ & $[-60^\circ,60^\circ]$ & $[\phi_{\min},\phi_{\max}]$ & $[-20^\circ,20^\circ]$ \\
Optimizer & Adam & Learning rate & $10^{-3}$ \\
Batch size & $128$ & Epochs & $100$ \\
Soft label variance & $\sigma_b^2=1$ & Loss weight & $\lambda=4$ \\
\hline
\end{tabular}
\end{table}

\subsection{Result Analysis}

We evaluate beam prediction using Top-$k$ angular accuracy and distance-based accuracy (DBA), which checks whether the predicted angular beam index lies within a beam-index neighborhood of the ground truth~\cite{charan2022multi}. Let $(i^\star,j^\star)$ and $(\hat i,\hat j)$ denote the ground-truth and predicted angular beam indices, respectively. DBA is defined as
\begin{equation}
\mathrm{DBA}
=
\frac{1}{N}
\sum_{n=1}^{N}
\mathbb{I}
\left(
\sqrt{
(\hat i_n-i_n^\star)^2
+
(\hat j_n-j_n^\star)^2
}
\leq \Delta
\right),
\end{equation}
where $\mathbb{I}(\cdot)$ is the indicator function and $\Delta=5$ in our experiments. Communication performance is further measured by the average beam loss (AvgL), median beam loss (MedL), and normalized spectral efficiency (SE).  

The proposed beam map learning method is compared with two ablation variants and two representative baselines: \textbf{Proposed w/o weighted MSE}, which removes the neighborhood-weighted MSE loss; \textbf{Proposed w/o soft label}, which uses one-hot beam labels; \textbf{CNN-based beam classification}, which formulates angular beam prediction as a multi-class classification problem and directly predicts the optimal angular beam index; and \textbf{Radar-assisted beam selection}, which estimates the dominant radar range-angle response and selects the nearest communication beam.

\begin{figure*}[h]
    \centering
    \includegraphics[width=0.86\textwidth]{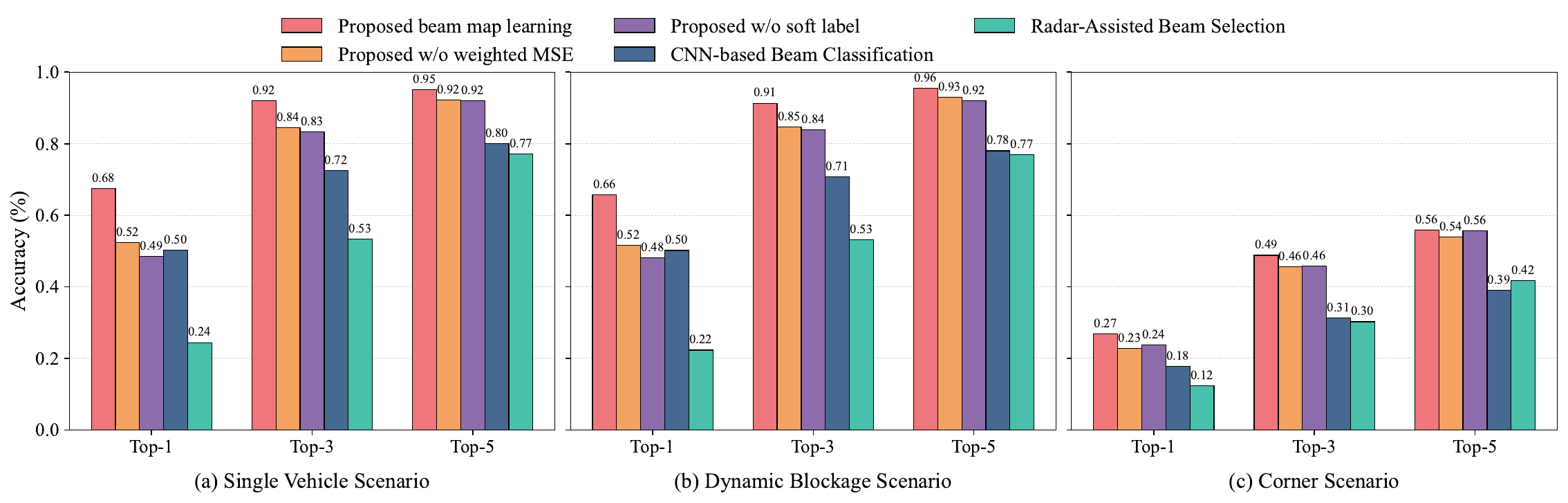}
    \caption{Top-$k$ beam prediction accuracy under different scenarios.}
    \label{fig:scene_topk_accuracy}
\end{figure*}

Fig.~\ref{fig:scene_topk_accuracy} illustrates the Top-$k$ accuracy across different scenarios, where the proposed framework consistently outperforms all baselines. The radar-assisted baseline yields poor accuracy due to severe physical mismatches in heterogeneous arrays. In the challenging corner NLoS scenario, the conventional CNN classification also suffers a severe performance collapse from overfitting. Conversely, by combining soft angular labels and the neighborhood-weighted MSE loss, the proposed method preserves physical spatial continuity, yielding consistent gains over the baselines.

\begin{table}[t]
\centering
\caption{Beam prediction performance under different vehicle mobility conditions.}
\label{tab:beam_prediction_results}
\scriptsize
\setlength{\tabcolsep}{2.5pt}
\renewcommand{\arraystretch}{0.95}
\begin{tabular}{llccccc}
\toprule
Speed & Metric
& Proposed
& w/o wMSE
& w/o soft
& CNN cls.
& Radar sel. \\
\midrule

\multirow{4}{*}{$<15$ m/s}
& Top-1$\uparrow$
& 0.543
& 0.448
& 0.399
& 0.429
& 0.212 \\
& DBA$\uparrow$
& 0.766
& 0.733
& 0.724
& 0.642
& 0.638 \\
& AvgL (dB)$\downarrow$
& 2.444
& 2.715
& 2.794
& 7.240
& 3.622 \\
& MedL (dB)$\downarrow$
& 0.200
& 0.345
& 0.474
& 0.789
& 0.991 \\
\midrule

\multirow{4}{*}{$\geq15$ m/s}
& Top-1$\uparrow$
& 0.520
& 0.386
& 0.404
& 0.342
& 0.174 \\
& DBA$\uparrow$
& 0.805
& 0.766
& 0.763
& 0.614
& 0.675 \\
& AvgL (dB)$\downarrow$
& 1.726
& 1.984
& 2.262
& 7.899
& 3.206 \\
& MedL (dB)$\downarrow$
& 0.207
& 0.360
& 0.361
& 1.528
& 0.813 \\
\midrule

\multirow{4}{*}{Overall}
& Top-1$\uparrow$
& \textbf{0.533}
& 0.423
& 0.401
& 0.394
& 0.197 \\
& DBA$\uparrow$
& \textbf{0.782}
& 0.747
& 0.740
& 0.631
& 0.653 \\
& AvgL (dB)$\downarrow$
& \textbf{2.156}
& 2.422
& 2.581
& 7.504
& 3.455 \\
& MedL (dB)$\downarrow$
& \textbf{0.202}
& 0.355
& 0.432
& 1.063
& 0.913 \\
\bottomrule
\end{tabular}
\vspace{-2mm}
\end{table}

Table~\ref{tab:beam_prediction_results} evaluates the framework's robustness across different mobility conditions. The proposed method consistently achieves the highest accuracy and lowest beam loss. Notably, performance remains stable across vehicle speeds, with the overall median loss of $0.202$~dB. This suggests that the learned radar-communication spatial correlation is robust to vehicular dynamics across the evaluated V2I mobility profiles.

\begin{figure}[t]
    \centering
    \includegraphics[width=0.78\linewidth]{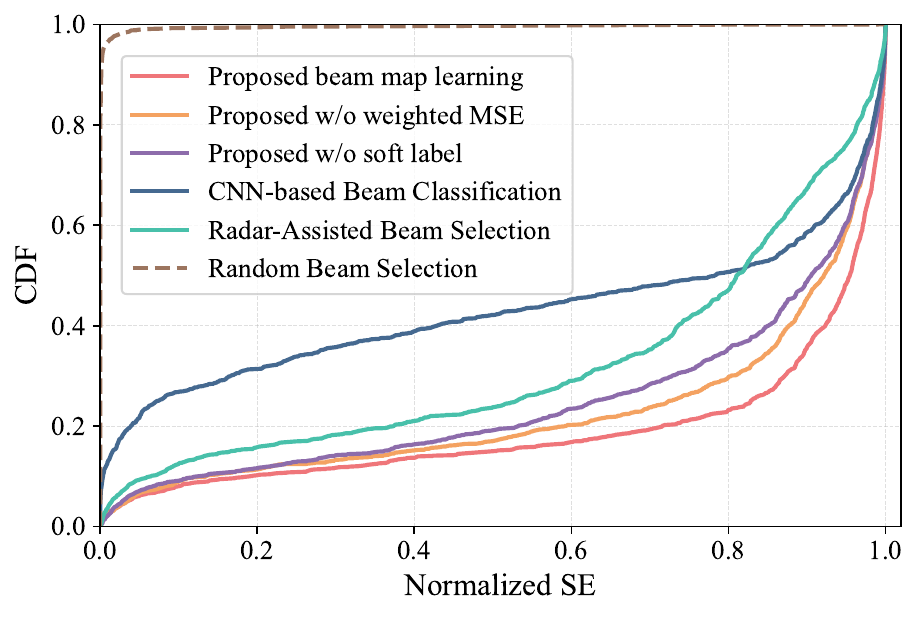}
    \caption{CDF of the SE achieved by different beam prediction methods.}
    \label{fig:normalized_se_cdf}
\end{figure}

Fig.~\ref{fig:normalized_se_cdf} shows the CDF of the normalized SE, defined as the ratio between the average SE achieved by the predicted beam and that achieved by the exhaustive-search optimal beam. The proposed method achieves the highest SE and approaches the optimal-beam performance for most samples. The ablation curves show that soft angular encoding provides the primary improvement, while neighborhood-weighted MSE yields further refinement. In contrast, the CNN classification struggles with hard label assignment errors, and pure radar-assisted selection is limited by geometric offsets. These results indicate that regressing a continuous beam map is more effective than discrete index matching in this setting.

\section{Conclusion}

This paper proposed a radar-aided near-field beam prediction framework for XL-MIMO V2I communications. The proposed method learns a radar-to-beam map from radar Bartlett spectra to communication angular beam map using Gaussian soft supervision and a lightweight encoder-decoder CNN. Evaluations on a synchronized Sionna RT-based radar-communication dataset demonstrate that the proposed method improves beam prediction accuracy and communication performance compared with hard beam classification and radar-assisted beam selection for XL-MIMO near-field V2I systems.




\bibliographystyle{IEEEtran}
\bibliography{refs}

@ARTICLE{10379539,
  author={Wang, Zhe and Zhang, Jiayi and Du, Hongyang and Niyato, Dusit and Cui, Shuguang and Ai, Bo and Debbah, Mérouane and Letaief, Khaled B. and Poor, H. Vincent},
  journal={IEEE Commun. Surv. Tutorials}, 
  title={A Tutorial on Extremely Large-Scale {MIMO} for 6{G}: Fundamentals, Signal Processing, and Applications}, 
  year={2024},
  volume={26},
  number={3},
  pages={1560-1605},
  month={Thirdquarter}}

@ARTICLE{cui2023nearfield,
  author={Cui, Mingyao and Wu, Zidong and Lu, Yu and Wei, Xiuhong and Dai, Linglong},
  journal={IEEE Commun. Mag.}, 
  title={Near-Field {MIMO} Communications for 6{G}: Fundamentals, Challenges, Potentials, and Future Directions}, 
  year={2023},
  volume={61},
  number={1},
  pages={40-46},
  month={Jan.}}

@ARTICLE{nie2025nearfield,
  author={Nie, Jiali and Cui, Yuanhao and Yang, Zhaohui and Yuan, Weijie and Jing, Xiaojun},
  journal={IEEE Trans. Mob. Comput.}, 
  title={Near-Field Beam Training for Extremely Large-Scale {MIMO} Based on Deep Learning}, 
  year={2025},
  volume={24},
  number={1},
  pages={352-362},
  month={Jan.}}

@ARTICLE{liu2022isac,
  author={Liu, Fan and Cui, Yuanhao and Masouros, Christos and Xu, Jie and Han, Tony Xiao and Eldar, Yonina C. and Buzzi, Stefano},
  journal={IEEE J. Sel. Areas Commun.}, 
  title={Integrated Sensing and Communications: Toward Dual-Functional Wireless Networks for 6{G} and Beyond}, 
  year={2022},
  volume={40},
  number={6},
  pages={1728-1767},
  month={Jun.}}

@ARTICLE{cui2024sensing,
  author={Cui, Yuanhao and Nie, Jiali and Cao, Xiaowen and Yu, Tiankuo and Zou, Jiaqi and Mu, Junsheng and Jing, Xiaojun},
  journal={IEEE J. Sel. Top. Signal Process.}, 
  title={Sensing-Assisted High Reliable Communication: A {Transformer}-Based Beamforming Approach}, 
  year={2024},
  volume={18},
  number={5},
  pages={782-795},
  month={Jul.}}

@ARTICLE{jiang2023lidar,
  author={Jiang, Shuaifeng and Charan, Gouranga and Alkhateeb, Ahmed},
  journal={IEEE Wireless Commun. Lett.}, 
  title={{LiDAR} Aided Future Beam Prediction in Real-World Millimeter Wave {V2I} Communications}, 
  year={2023},
  volume={12},
  number={2},
  pages={212-216},
  month={Feb.}}

@ARTICLE{charan2025sensing,
  author={Charan, Gouranga and Alkhateeb, Ahmed},
  journal={IEEE Open J. Commun. Soc.}, 
  title={Sensing-Aided 6{G} Drone Communications: Real-World Datasets and Demonstration}, 
  year={2025},
  volume={6},
  number={},
  pages={8745-8774},
  month={Sep.}}

@ARTICLE{liu2020radar,
  author={Liu, Fan and Yuan, Weijie and Masouros, Christos and Yuan, Jinhong},
  journal={IEEE Trans. Wireless Commun.}, 
  title={Radar-Assisted Predictive Beamforming for Vehicular Links: Communication Served by Sensing}, 
  year={2020},
  volume={19},
  number={11},
  pages={7704-7719},
  month={Nov.}}

@INPROCEEDINGS{demirhan2022radar,
  author={Demirhan, Umut and Alkhateeb, Ahmed},
  booktitle={2022 IEEE Wireless Communications and Networking Conference (WCNC)}, 
  title={Radar Aided 6{G} Beam Prediction: Deep Learning Algorithms and Real-World Demonstration}, 
  year={2022},
  volume={},
  number={},
  pages={2655-2660},
  month={Apr.}}

@ARTICLE{demirhan2026radartracking,
  author={Demirhan, Umut and Alkhateeb, Ahmed},
  journal={IEEE Trans. Commun.}, 
  title={Radar-Aided Beam Prediction and Tracking: Will It Work in the Real World?}, 
  year={2026},
  volume={74},
  number={},
  pages={4336-4352},
  }

@ARTICLE{ali2020passive,
  author={Ali, Anum and González-Prelcic, Nuria and Ghosh, Amitava},
  journal={IEEE Trans. Veh. Technol.}, 
  title={Passive Radar at the Roadside Unit to Configure Millimeter Wave Vehicle-to-Infrastructure Links}, 
  year={2020},
  volume={69},
  number={12},
  pages={14903-14917},
  month={Dec.}}

@ARTICLE{graff2023deep,
  author={Graff, Andrew and Chen, Yun and González-Prelcic, Nuria and Shimizu, Takayuki},
  journal={IEEE Trans. Veh. Technol.}, 
  title={Deep Learning-Based Link Configuration for Radar-Aided Multiuser mmWave Vehicle-to-Infrastructure Communication}, 
  year={2023},
  volume={72},
  number={6},
  pages={7454-7468},
  month={Jun.}}

@ARTICLE{charan2022multi,
  title={Multi-modal beam prediction challenge 2022: Towards generalization},
  author={Charan, Gouranga and Demirhan, Umut and Morais, Joao and Behboodi, Arash and Pezeshki, Hamed and Alkhateeb, Ahmed},
  journal={arXiv preprint arXiv:2209.07519},
  year={2022}
}

\end{document}